\documentclass[preprint,5p]{elsarticle}
\usepackage{amssymb}
\usepackage{graphicx}
\journal{Nuclear Instruments and Methods in Physics Research A}

\begin{document}
\begin{frontmatter}

\title{Large-scale Gadolinium-doped Water \v{C}erenkov Detector for Non-Proliferation}
\author[llnl,davis]{M. Sweany}
\author[llnl]{A. Bernstein}
\author[llnl]{N.S. Bowden}
\author[llnl]{S. Dazeley}
\author[llnl]{G. Keefer}
\author[davis]{R. Svoboda}
\author[davis]{M. Tripathi}

\address[llnl]{Lawrence Livermore National Laboratory, Livermore, CA 94550, USA}
\address[davis]{Department of Physics, University of California, Davis, CA 95616, USA}
\date{\today}

\begin{abstract}
Fission events from Special Nuclear Material (SNM), such as highly enriched uranium or plutonium, can produce simultaneous emission of multiple neutrons and high energy gamma-rays.  The observation of time correlations between any of these particles is a significant indicator of the presence of fissionable material. Cosmogenic processes can also mimic these types of correlated signals.  However, if the background is sufficiently low and fully characterized, significant changes in the correlated event rate in the presence of a target of interest constitutes a robust signature of the presence of SNM.  Since fission emissions are isotropic, adequate sensitivity to these multiplicities requires a high efficiency detector with a large solid angle with respect to the target.  Water \v{C}erenkov detectors are a cost-effective choice when large solid angle coverage is required.  In order to characterize the neutron detection performance of large-scale water \v{C}erenkov detectors, we have designed and built a 3.5 kL water \v{C}erenkov-based gamma-ray and neutron detector, and modeled the detector response in Geant4 \cite{geant4}.  We report the position-dependent neutron detection efficiency and energy response of the detector, as well as the basic characteristics of the simulation. 
\end{abstract}
\end{frontmatter}

\section{Introduction}
Legitimate cross border trade involves the transport of an enormous number of cargo containers. In order to verify that these containers are not transporting SNM without impeding legitimate trade, there is a need for fast, highly efficient, and large detectors that are relatively inexpensive. Such detectors must produce consistent yet distinct responses to both SNM and background, so that their effectiveness is not reduced by false positive or negative detections. They also need to have limited sensitivity to background radiation, such as cosmic ray induced background or Naturally Occurring Radioactive Material (NORM) present in certain legitimate forms of cargo. Both of these may contribute to false positives or reduce sensitivity to real SNM.

SNM can either spontaneously fission or be induced to do so by an external source of gamma rays or neutrons. Since cargo containers are large, they can contain a significant amount of shielding. The fission emissions most likely to penetrate the container and interact with a detector are neutrons or high energy (greater than 3 MeV) gamma-rays. We propose that a water \v{C}erenkov detector doped with a neutron capturing agent (such as GdCl$_3$ salt) would be ideal for this application.  Such a detector, sensitive to short timescale correlations between events, has a number of advantages; it is relatively inexpensive, nonflammable and noncombustible, environmentally safe, and easy to operate.  

\v{C}erenkov detectors do not produce correlated signals from single fast neutrons in the same way organic scintillator does: in scintillator, fast neutrons are capable of producing a correlated signal via proton recoil followed by neutron capture.  Water \v{C}erenkov detectors only use the thermal neutron capture and prompt gamma-ray signals.

Thermal neutron capture on natural Gadolinium has an extremely high cross section (49,000 barns). On capture, a gamma-ray shower with energies adding to approximately 8 MeV is produced, and \v{C}erenkov radiation produced by the resulting Compton scatters is detectable by ordinary PMTs. SNO and Super-Kamiokande have shown that the \v{C}erenkov process can generate enough photons in water to detect neutron captures or gamma-rays with an energy of approximately 3 MeV or greater, so long as the photocathode coverage is high ($\sim$40\%) \cite{sno,SK}.  Our group has since demonstrated the viability of this technique above ground, operating 250 liter water based neutron detector with a photocathode coverage of 10\%; reflective detector walls made a lower photocathode coverage possible \cite{dazeley}. 

\section{Detector Description}
\begin{figure}\begin{center}
\includegraphics[width=3.5in]{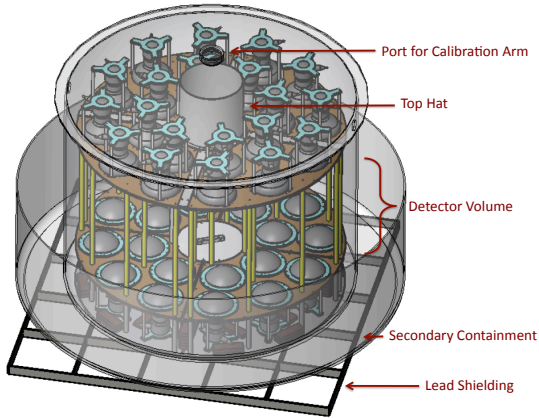}
\caption{The detector schematic showing the port and top hat for calibration arm insertion.  The detector volume is in-between two PMT arrays supported by acrylic rods, and the detector sits inside a secondary containment vessel to protect against accidental leaks.  Lead bricks shield against background gamma radiation from the floor.}
\label{fig:WNDfinished}
\end{center}\end{figure}

\begin{figure}[!t]
\begin{center}
\includegraphics[width=3.5in]{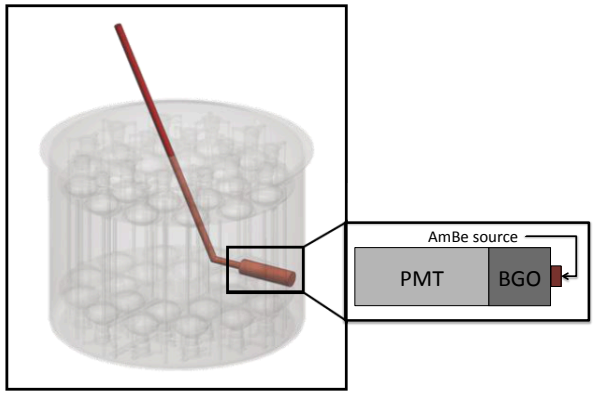}
\caption{The neutron calibration arm:  the PMT reads out gamma radiation coming from the AmBe source.}
\label{fig:wndCal}
\end{center}\end{figure}

In order to measure the neutron detection efficiency and maximize the performance of a large-scale detector, we have constructed a 3.5 kL water \v{C}erenkov-based neutron detector.  The detector consists of a cylindrical polyethylene tank with two PMT arrays arranged on the bottom and top, each with twenty Hamamatsu R7081 10 inch PMTs: although the current generation of R7081 PMTs are high quantum efficiency, the model used here does not have that improvement. The total photocathode coverage is approximately 19\%.  The tank is approximately 1.5 meter high, with a 2 meter diameter.  The detector sits atop a layer of lead bricks to shield against background gamma radiation from the floor. Figure \ref{fig:WNDfinished} shows a schematic of the detector: the white cylindrical opening in the center, called the top hat, allows for  deployment of calibration sources at several locations inside the detector through a 10 cm diameter portal in the lid.  The calibration source sits at the end of a long arm (the calibration arm), shown in Figure \ref{fig:wndCal} and discussed further below.

\begin{figure}[!t]
\begin{center}
\includegraphics[width=1.71in]{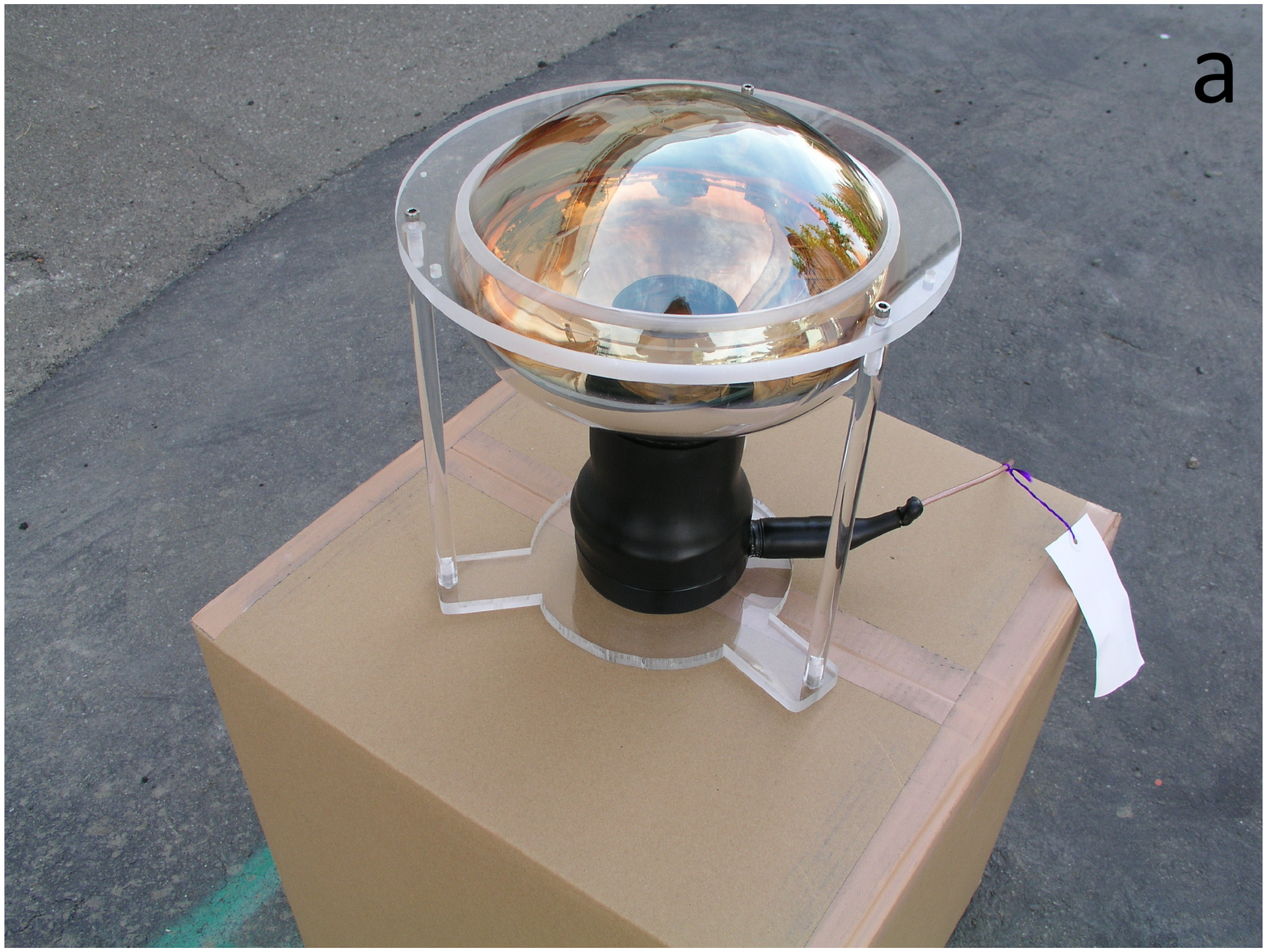}
\includegraphics[width=1.71in]{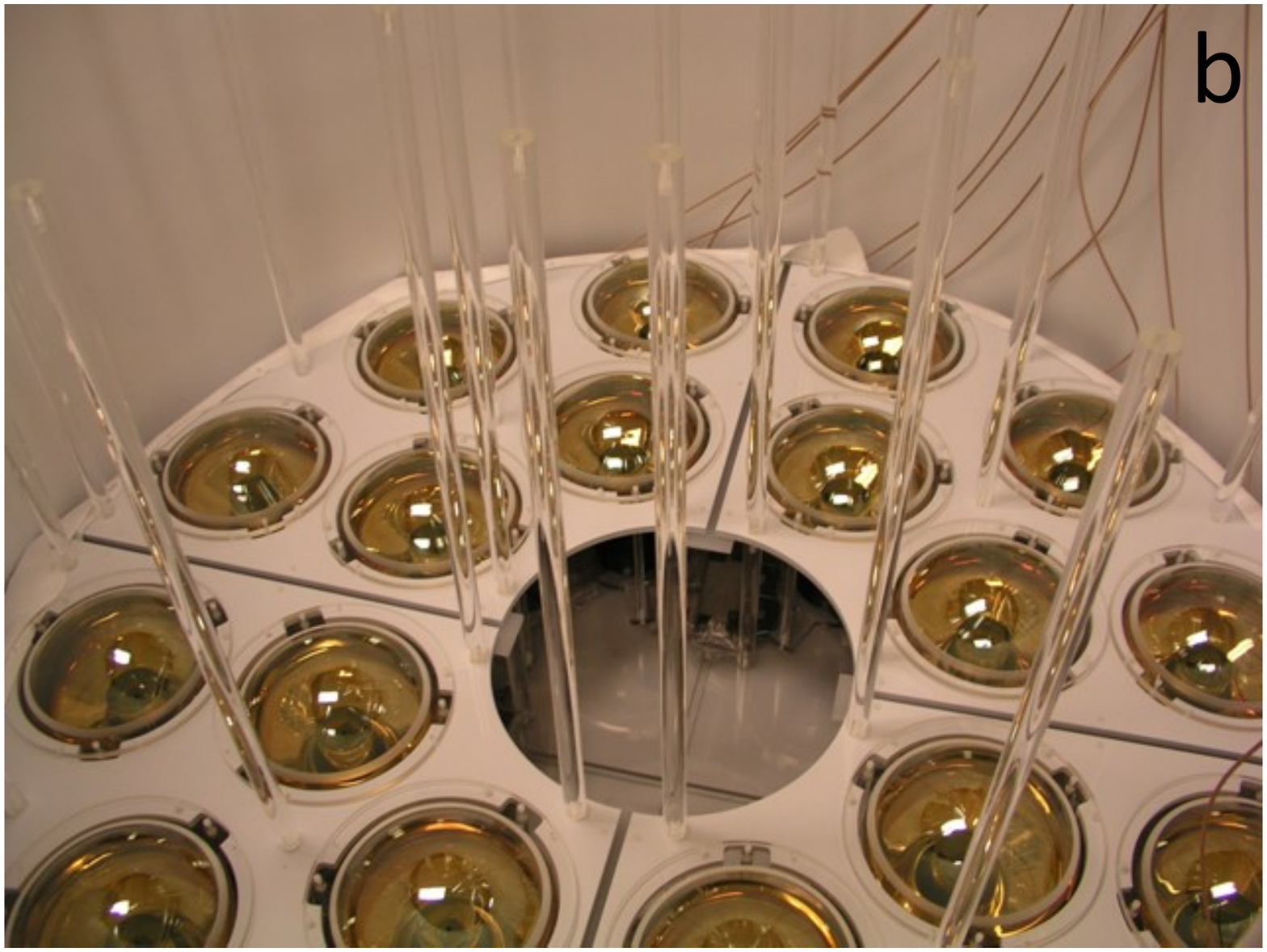}
\caption{A Hamamatsu R7081 PMT with factory water proof potting (a), and the bottom 20 PMTs installed in the bottom detector array (b).}
\label{fig:waterPmt}
\end{center}\end{figure}

It has been demonstrated that GdCl$_3$ doped water is not compatible with stainless steel, contact with which results in reduced water clarity \cite{coleman}.   All detector components in contact with the water are therefore constructed of either plastic or glass.  The PMTs are mounted on a frame constructed out of acrylic and white polypropylene that sits inside the tank, shown in Figure \ref{fig:waterPmt}a.   Each detector quadrant contains ten PMTs, five each on the top and bottom, with the top PMT array supported by acrylic rods.  Because the frame structure and PMTs are buoyant, stainless steel bars (sealed inside polypropylene bags) are used as ballast.  Figure \ref{fig:waterPmt}b, shows the bottom four quadrants with the support rods in place.  To increase light collection, the detector wall in between the two PMT arrays is lined with UV reflecting Teflon.

The detector is filled from the bottom using a water purification system capable of obtaining ultra pure deionized water (resistivity greater than 17 M$\Omega \cdot$cm).  Municipal water is passed through three large de-ionizing (DI) resin bottles and is then circulated through a purification system purchased from South Coast Water Inc., consisting of a UV sterilizer, 5 micron and 0.22 micron filters, and an additional DI unit.  Each unit in the system can be by-passed through a series of valves.  After purification, the DI unit is by-passed and the water is doped with GdCl$_3$.  Any particulate matter and remaining biological contamination from the GdCl$_3$ is eliminated by continued re-circulation through the filters and UV sterilizer.  Finally, the doped water is sent to the detector inlet.

\subsection{Data Acquisition}
The DAQ and trigger electronics are mostly commercial VME and NIM modules.  The only modules designed and fabricated in-house are eight-channel signal pick-off modules.  The pick-off modules are needed to separate the fast PMT signal from the high voltage bias, which is carried on the same cable coming out of the PMT. 

From there, the PMT signals are passed through a Mini-Circuits 15542 BBLP-39+, 23 MHz low-pass filter in order to stretch out the signal, then amplified and digitized using CAEN V975 amplifiers and 200 MHz Struck SIS3320 digitizers.  There are two triggers for the system: an internally generated trigger for physics events in the water and an external trigger used for calibration.  The internal trigger is formed from a four-fold coincidence among a group of 16 bottom PMT channels.  The PMT signals are fed to a CAEN V814 discriminator, set to trigger at $\sim$1 photo-electron.  A CAEN V1495 FPGA registers the 4-fold coincidences and issues the trigger to the waveform digitizers.  

\section{Calibration and Detection Efficiency}

Calibration of the detector is needed to equalize the PMT gains and to set an approximate energy scale for physics events.  We have three artificial sources: an LED for PMT gain, a $^{252}$Cf fission source for neutrons, and a tagged neutron source to measure the neutron detection performance on an event-by-event basis.   A polyethylene calibration arm was constructed in collaboration with a group at Harvey Mudd College to deploy the neutron sources at multiple positions inside the detector.

The tagged neutron source consists of an americium beryllium (AmBe) source and a Scionix type 51B51/2M-E1-BGO crystal and PMT deployed together inside the calibration arm.  AmBe sources emit a neutron in coincidence with a 4.4 MeV gamma-ray.  Detection of this gamma-ray by the BGO detector forms a "tag", indicating that a neutron has been emitted from the source.  A rendering of the crystal, PMT, and AmBe source inside the polyethylene arm is shown in Figure \ref{fig:wndCal}.  The arm was designed so that it can reach several positions inside the detector, making position dependent neutron detection efficiency measurements possible.

\subsection{ $^{252}$Cf Calibration}
Three datasets were taken with a $^{252}$Cf source positioned at various locations outside the detector: seven inches from the detector wall, one meter from the detector wall, and two meters from the detector wall.  The first position is used to calibrate the response from Monte Carlo and to establish the quality of event-level cuts on the data.  The one and two meter positions are used to determine how the event-level cuts perform with distant sources.

\begin{figure}[!t]
\centering
\includegraphics[width=3.5in]{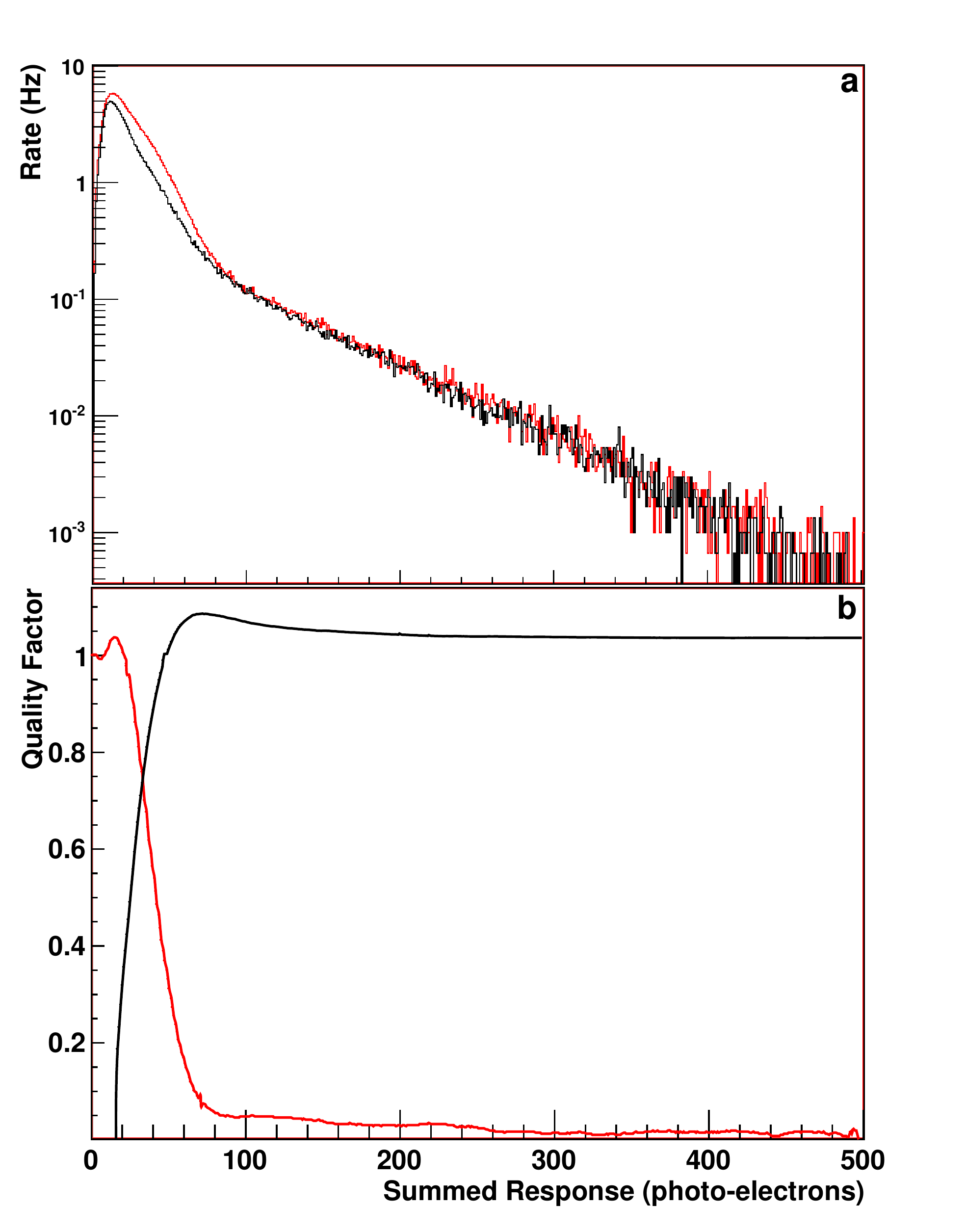}
\caption{In (a), the correlated (red) and uncorrelated (black) summed response with the $^{252}$Cf source seven inches from the detector edge: the high-energy tail is due to cosmic-ray induced muons and gamma-rays.  The charge quality factor for left (red) and right (black) cuts is shown in (b), where the maximum in the quality factor is used to establish the cut values.  }
\label{fig:chargeQF}
\end{figure}

\begin{figure}[!t]
\centering
\includegraphics[width=3.5in]{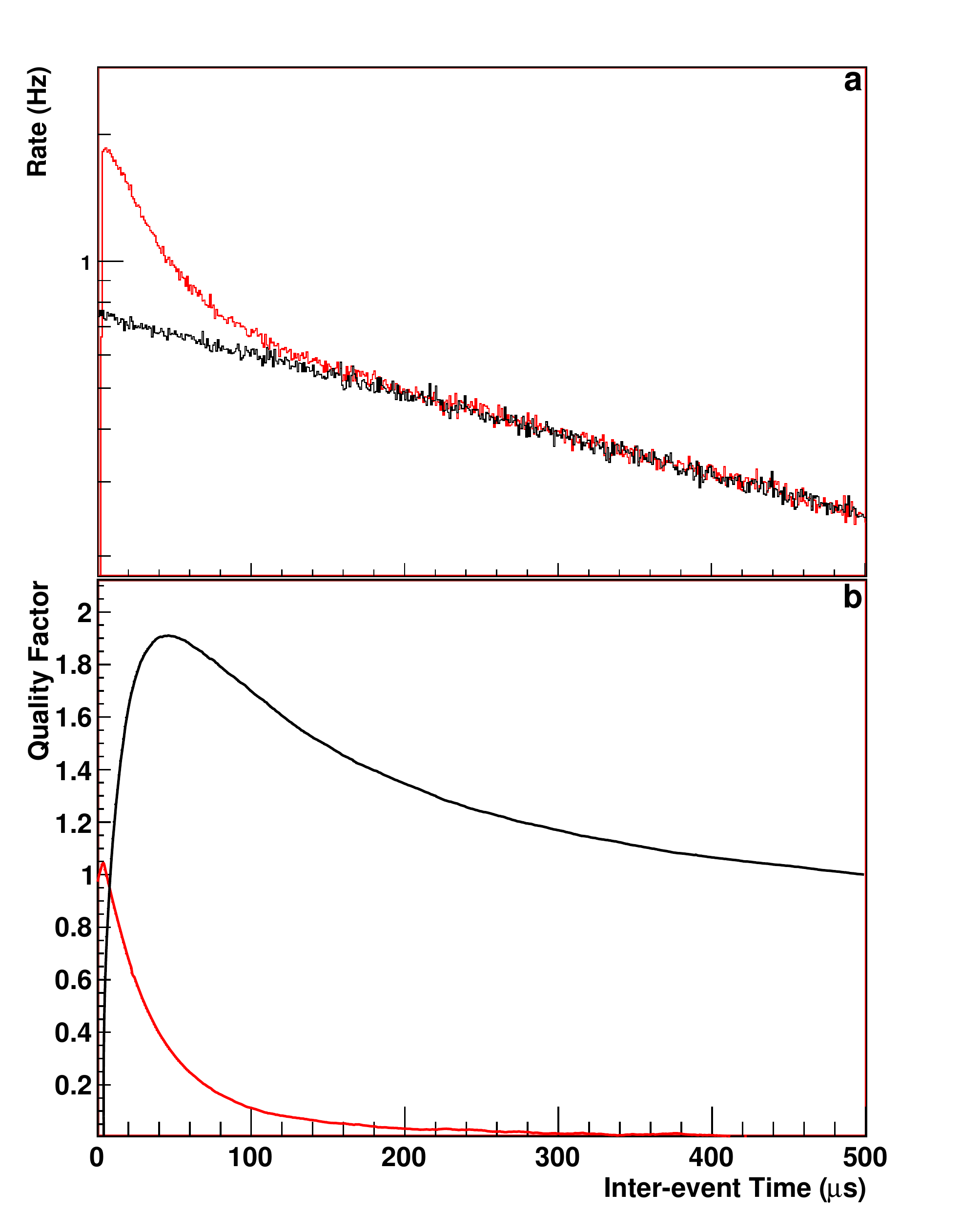}
\caption{In (a), the inter-event time for events passing the charge cut (red), and the fit to the uncorrelated background (black).  The time quality factor for left (red) and right (black) cuts is shown in (b), where the maximum in the quality factor is used to establish the cut values.}
\label{fig:timeQF}
\end{figure}

Since high energy events, such as muons traversing the detector tend to saturate the response of the PMTs, we first screen our data to remove all events that contain at least one saturated PMT.  Event level cuts are then applied to select neutrons in the detector.  Each is determined by maximizing a Quality Factor, $Q$, defined as the significance after the cut is applied divided by the significance before the cut is applied.  For a given parameter distribution divided into n independent bins, and for an analysis cut at the jth bin, the quality factor approaching from the left, or lower bound, is

\[
Q^{\mathrm{L}}_j  = \left(\frac{\sum\limits_{i=0}^{j} \mathrm{On}_i - \sum\limits_{i=0}^{j}\mathrm{Off}_i}{\sqrt{\sum\limits_{i=0}^{j}\mathrm{Off}_i}}\right)\left(\frac{\sqrt{\sum\limits_{i=0}^{n}\mathrm{Off}_i}}{\sum\limits_{i=0}^{n} \mathrm{On}_i - \sum\limits_{i=0}^{n}\mathrm{Off}_i}\right),
\]

where $\mathrm{On}$ and $\mathrm{Off}$ represent the value of the parameter with and without the neutron source present, respectively.  $Q$ is defined to have a value equal to 1 when no cut is applied, i.e. when j equals n.  Once a lower bound cut value has been determined, we then apply the same formula in reverse, or from the right, to determine the best jth bin for an upper bound cut:

\[
Q^{\mathrm{R}}_j = \left(\frac{\sum\limits_{i=n}^{j} \mathrm{On}_i - \sum\limits_{i=n}^{j}\mathrm{Off}_i}{\sqrt{\sum\limits_{i=n}^{j}\mathrm{Off}_i}}\right)\left(\frac{\sqrt{\sum\limits_{i=0}^{n}\mathrm{Off}_i}}{\sum\limits_{i=0}^{n} \mathrm{On}_i - \sum\limits_{i=0}^{n}\mathrm{Off}_i}\right).
\]

The quality analysis for charge is shown in Figure \ref{fig:chargeQF}.  Because the fission source increases the number of uncorrelated gamma-rays interacting in the detector, using the on source uncorrelated background as the no source present histogram for the quality analysis provides better rejection of gamma energies.   In Figure \ref{fig:chargeQF}a, the correlated on source data is shown in red and the uncorrelated on source is shown in black.  Figure \ref{fig:chargeQF}b shows $Q$ for the left (pink) and right (blue) cuts.  The resulting cuts are applied to both the current event and the previous event to establish neutron-neutron pair events in the detector.

Since the presence of a neutron source results in an increase in the uncorrelated trigger rate, a comparison of the inter-event time distribution with and without the source is not appropriate.  Instead, the random trigger rate in the on source data is fit to an exponential, and a histogram  filled randomly from the fit values serves as the no source present histogram.  The results shown in Figure \ref{fig:timeQF} with the on source inter-event time, as well as the quality factor.  

In order to pick out neutron-neutron events, we cut on both the current event and the previous event charge.  Finally, we reject events in which the time difference between the current event and the last muon is less than 46 $\mu$s, which acts as a muon veto; the value is chosen as the same maximum inter-event time value allowed.  A muon veto cut of 46 $\mu$s rejects 9.9\% of the correlated background and increases the dead-time of the detector by 2.5\%.  Although there is modest improvement in rejecting muon-induced backgrounds, the majority of cosmic backgrounds appear to result from muons in the vicinity of the detector but not traversing it.  The final cut values for all four event-level cuts are shown in Table \ref{tab:cuts}.

\begin{table}[!t]
\centering
\begin{tabular}{lll}
Parameter		&Left Cut				&Right Cut	\\
\hline
Current Charge	& 16 	pe				& 72	pe		\\
Previous Charge 	& 16	pe				& 72	pe		\\
Inter-event Time	& 4 $\mu$s			& 46	$\mu$s	\\
Muon Veto 		&\textgreater46 $\mu$s	& N/A		 \\
\end{tabular}
\caption{Analysis cuts obtained by maximizing the signal significance between the background data run and $^{252}$Cf data run with the source seven inches from the detector.}
\label{tab:cuts}
\end{table}

\begin{table}[!t]
\centering
\begin{tabular}{lll}
Event Rate	(Hz)	&$^{252}$Cf Source	& No Source	\\
\hline
Raw  			& 2390 			& 1770 		\\
Singles 			& 920 			& 468 		\\
Doubles 			& 55.0 			& 13.7 		\\
\end{tabular}
\caption{Average event rates over 20 seconds for $^{252}$Cf data run with the source seven inches from the detector, compared with no source present.  The singles rate has only the energy cut and a muon veto applied, and the doubles rate has all cuts applied. }
\label{tab:rates}
\end{table}

Table \ref{tab:rates} shows the event rates averaged over 20 seconds for the data run with the $^{252}$Cf source seven inches from the detector edge compared to no source present.  At this position, the solid angle coverage is 28\% of 4$\pi$.  The singles rate, or the rate of single neutron events in the detector, is the rate after both an energy cut on the current event and the muon veto cut has been applied.  The doubles rate, or the rate of neutron-neutron pair events, is the rate after all cuts in Table \ref{tab:cuts} have been applied.

\subsection{Neutron Detection Efficiency}
\begin{figure}[!t]
\centering
\includegraphics[width=3.5in]{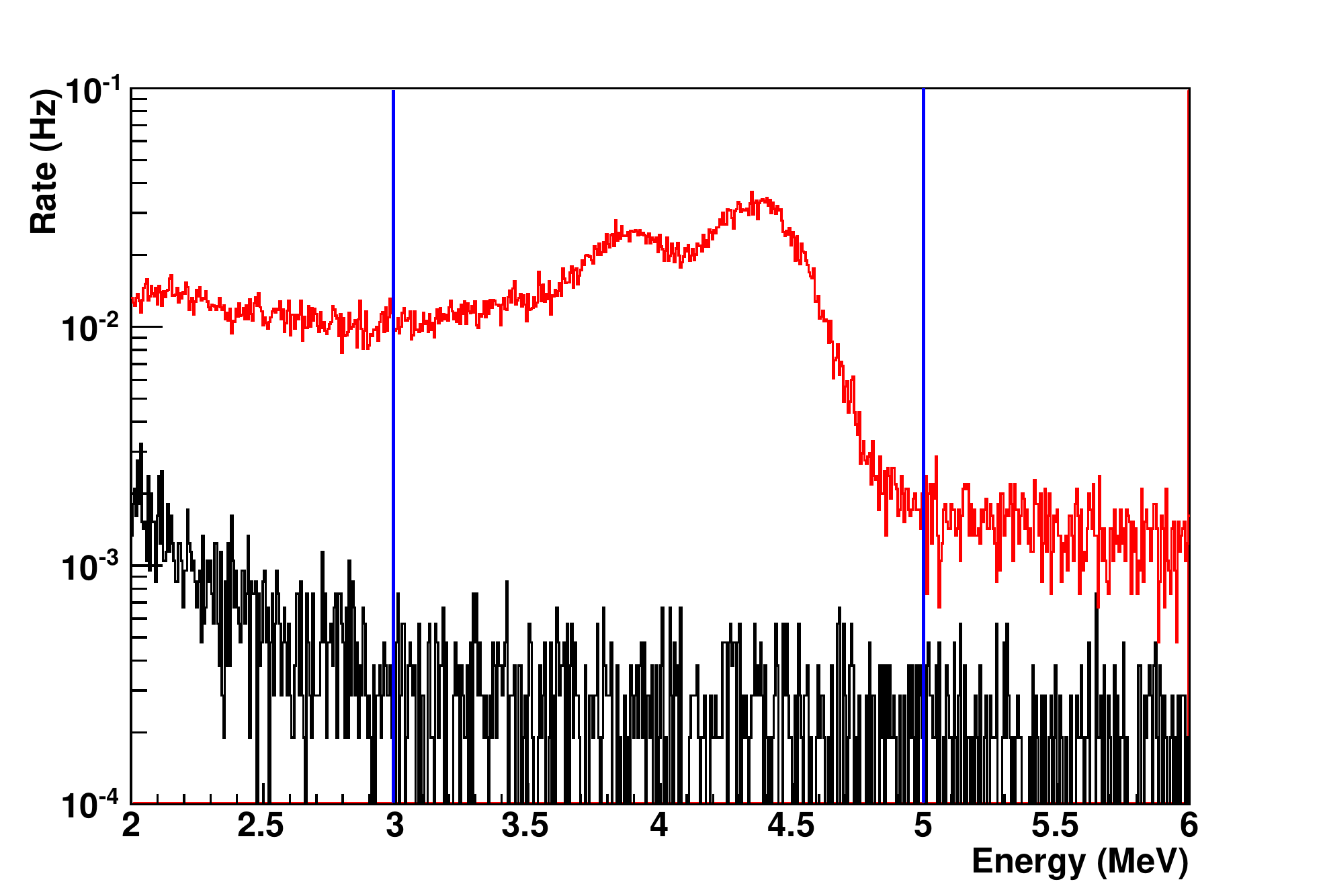}
\caption{The charge from the BGO crystal with (red) and without (black) an AmBe source present.  The energy cuts around the 4.4 MeV gammas and two escape peaks are shown in blue.}
\label{fig:BGO}
\end{figure}

\begin{figure}[!t]
\centering
\includegraphics[width=3.5in]{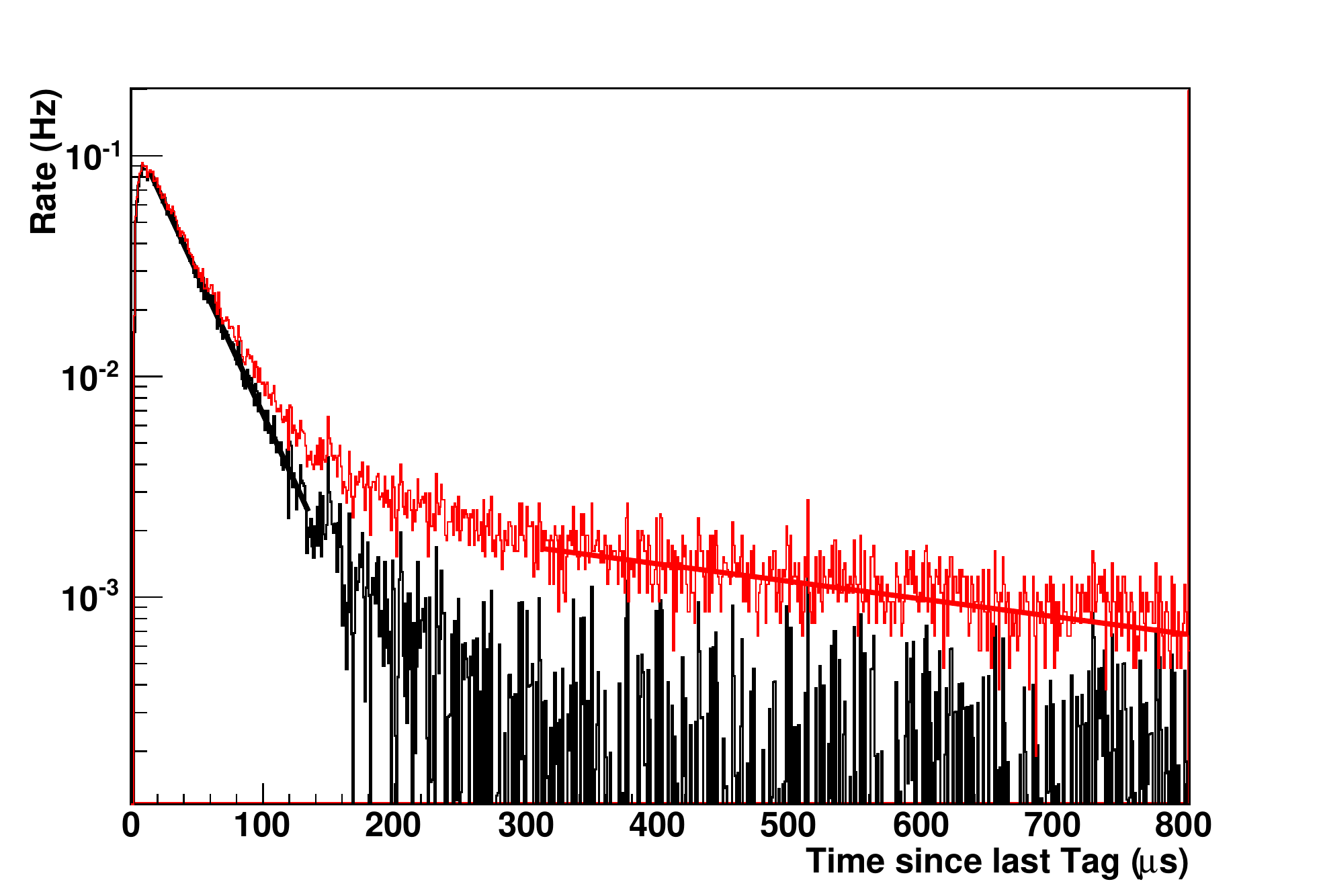}
\caption{The time between tag and detector triggers for all events (red).  In the 300-800 $\mu$s range are the uncorrelated accidentals.  The correlated events in the 0-100 $\mu$s range, with the uncorrelated background subtracted (black), indicate a capture time of 35 $\mu$s.}
\label{fig:AmBeTime}
\end{figure}

The neutron detection efficiency is determined using the calibration arm described above.  The AmBe gamma-ray spectrum from the BGO crystal in Figure \ref{fig:BGO} shows the primary 4.4 MeV gamma-ray peak.  The smaller peaks at 3.9 MeV and 3.4 MeV are also due to 4.4 MeV gamma-rays, where pair production and subsequent positron annihilation results in either one or two 511 keV gamma-rays escaping from the crystal.  To maximize the signal to background ratio of our tag, we select events in the range 3 MeV to 5 MeV, where the background gamma-ray rate in the crystal is very small.  Figure \ref{fig:AmBeTime} shows the timing distribution of delayed detector events after our selection of AmBe tags.  This inter-event time distribution can be well parameterized by a sum of two exponentials, representing a correlated and uncorrelated set of events.  The correlated set represents neutron capture events due to the AmBe source in our detector.  The mean capture time is 35 $\mu$s.  This is consistent with the expected capture time of thermal neutrons in water doped with $0.1\%$ gadolinium \cite{dazeley,apollonio,boehm,piepke}.  The number of neutron captures detected can be estimated by subtracting the exponential fit to the un-correlated events.  The efficiency is taken to be the integral of the correlated events, the black curve in Figure \ref{fig:AmBeTime}, divided by the total number of neutron tags (after accounting for the crystal's background rate), or the red curve in Figure \ref{fig:AmBeTime}.


\begin{figure}[!t]
\centering
\includegraphics[width=3.5in]{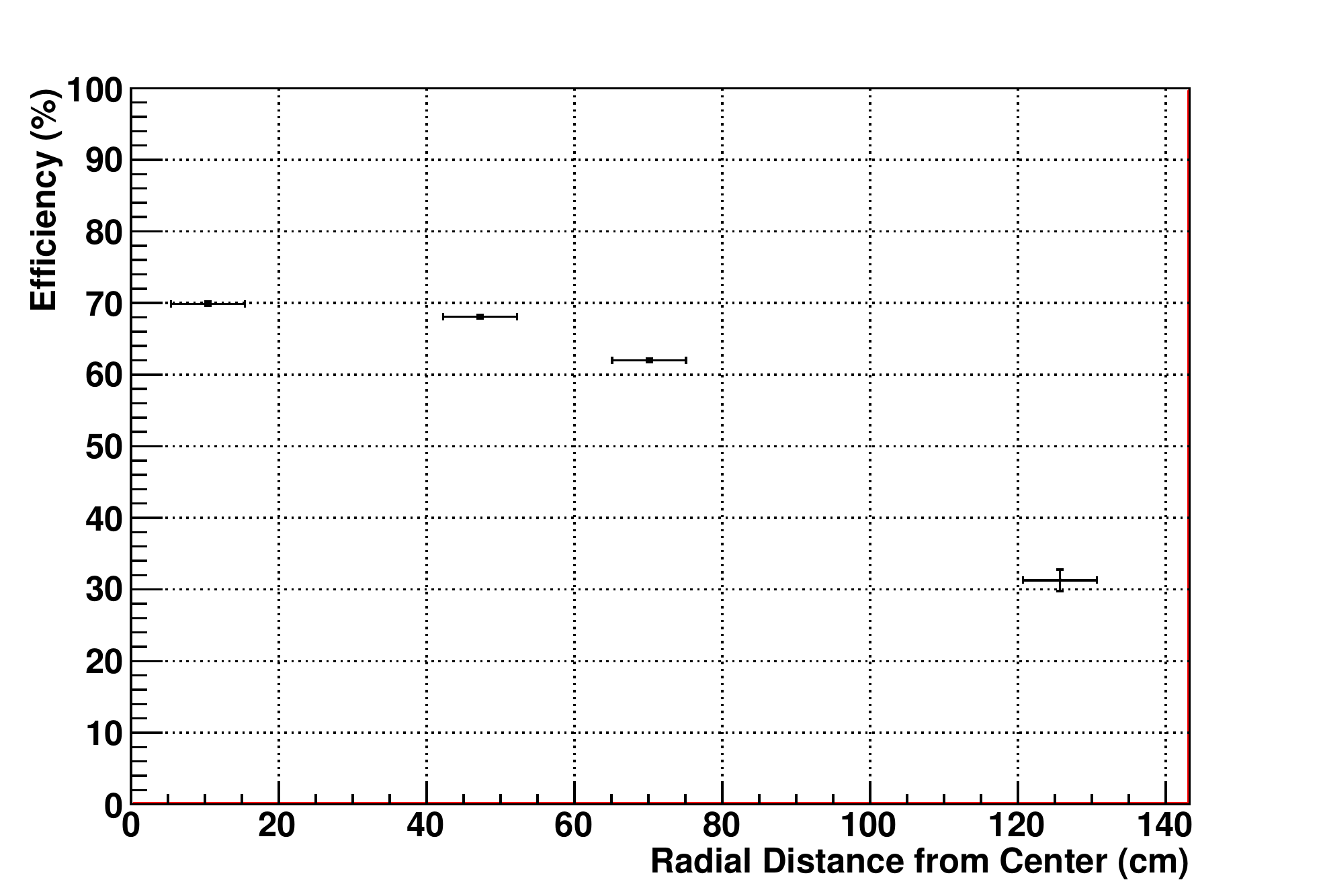}
\caption{The neutron detection efficiency verses the AmBe source position.  For the position outside the detector (125 cm), the efficiency has been scaled by the solid angle coverage of the detector.}
\label{fig:eff}
\end{figure}

Three AmBe data sets were taken inside the detector to determine the efficiency and energy response at various radii.  We assume that the detector is radially symmetric. One dataset was taken outside the detector, for which the efficiency is multiplied by the fraction of solid angle calculated from the source position.  The position-dependent efficiency is shown in Figure \ref{fig:eff}; it ranges from 69.9\% at the center of the detector to 31.3\% outside the detector.  The efficiency drop is expected near the detector edge, as neutrons can leave before being captured and many of the neutron capture gamma-rays escape the detector before interacting. 

\subsection{Monte Carlo Response}
A full Monte Carlo was written in Geant4 to model our detector.  Our objective is to utilize the Monte Carlo, tuned to reproduce the response of this detector, to determine the best design for a larger-scale detector.  

The output of the Monte Carlo is the wavelength of photon hits on the PMT surfaces.  In analysis, an energy-dependent quantum efficiency is applied, and a variable single photo-electron response is applied to each photo-electron.  The trigger is modeled by requiring at least four PMT signals to be greater than a given threshold.  The detector data is then scaled by the PE to ADC unit value determined from single photo-electron calibrations.

There are three optical properties that can be tuned to reproduce the response in data: the wall reflectivity, the attenuation length of water, and, to a limited extent, the PMT quantum efficiency.  The attenuation length of water in our detector was likely adversely affected by UV stabilizers in the polyethylene tank.  The attenuation length from \cite{quickenden, sogandares, pope} scaled down to an approximately 10 meter maximum gave results consistent with our data.  The quantum efficiency is taken from Hamamatsu specifications, and scaled down to account for losses in collection efficiency from stray magnetic fields: the total efficiency peaks at 22\%.  Finally, the wall reflectivity has a total reflectivity of 90\% with a 5\% specular component.

\begin{figure}[!t]
\centering
\includegraphics[width=3.5in]{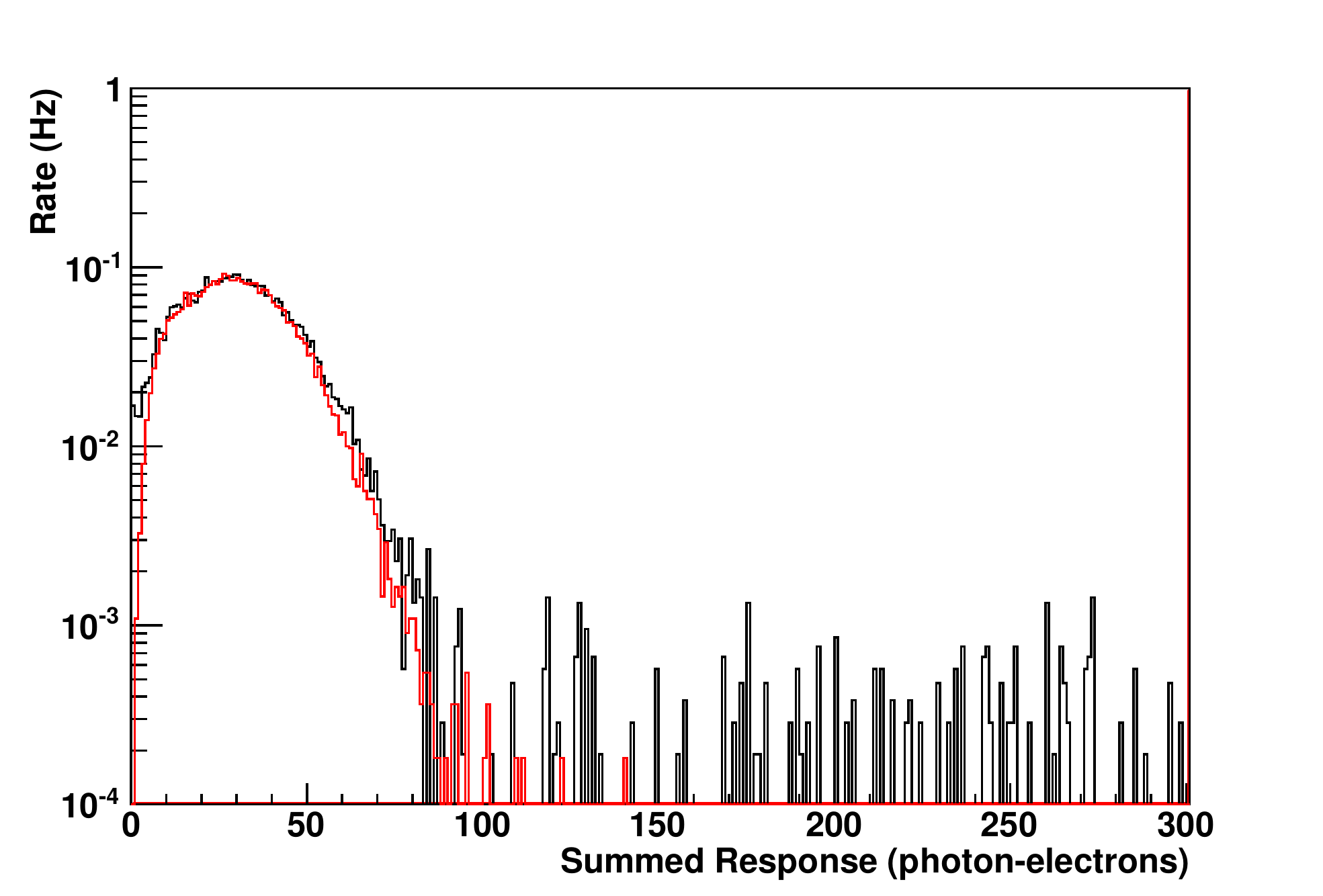}
\caption{The pure neutron spectrum obtained from the tagged-calibration arm positioned inside the detector (black) compared with the MC spectrum from 5 MeV neutrons generated at the center of the detector (red).}
\label{fig:AmBeMC}
\end{figure}

\begin{figure}[!t]
\centering
\includegraphics[width=3.5in]{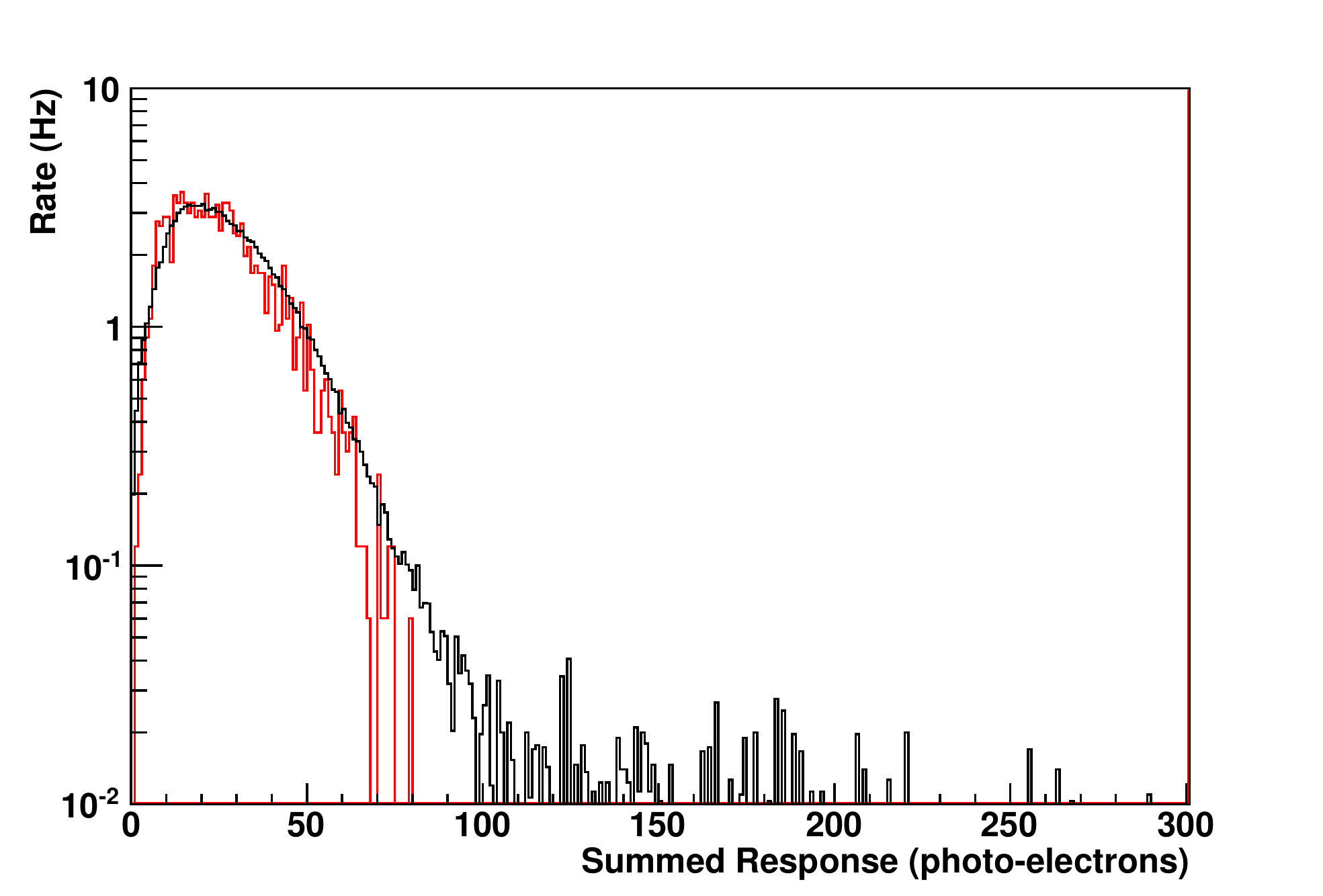}
\caption{The pure neutron spectrum obtained from a $^{252}$Cf source positioned 7 inches from the detector edge (black) compared with the MC spectrum from 5 MeV neutrons, 7 inches from the detector edge.}
\label{fig:CfMC}
\end{figure}

Good agreement between data and Monte Carlo has been obtained with both AmBe source data inside the detector (Figure \ref{fig:AmBeMC}) and the $^{252}$Cf source data outside the detector (Figure \ref{fig:CfMC}).  For the AmBe comparison, the calibration arm is included in the Monte Carlo.

\section{Discussion and Conclusions}
\begin{figure}[!t]
\centering
\includegraphics[width=3.5in]{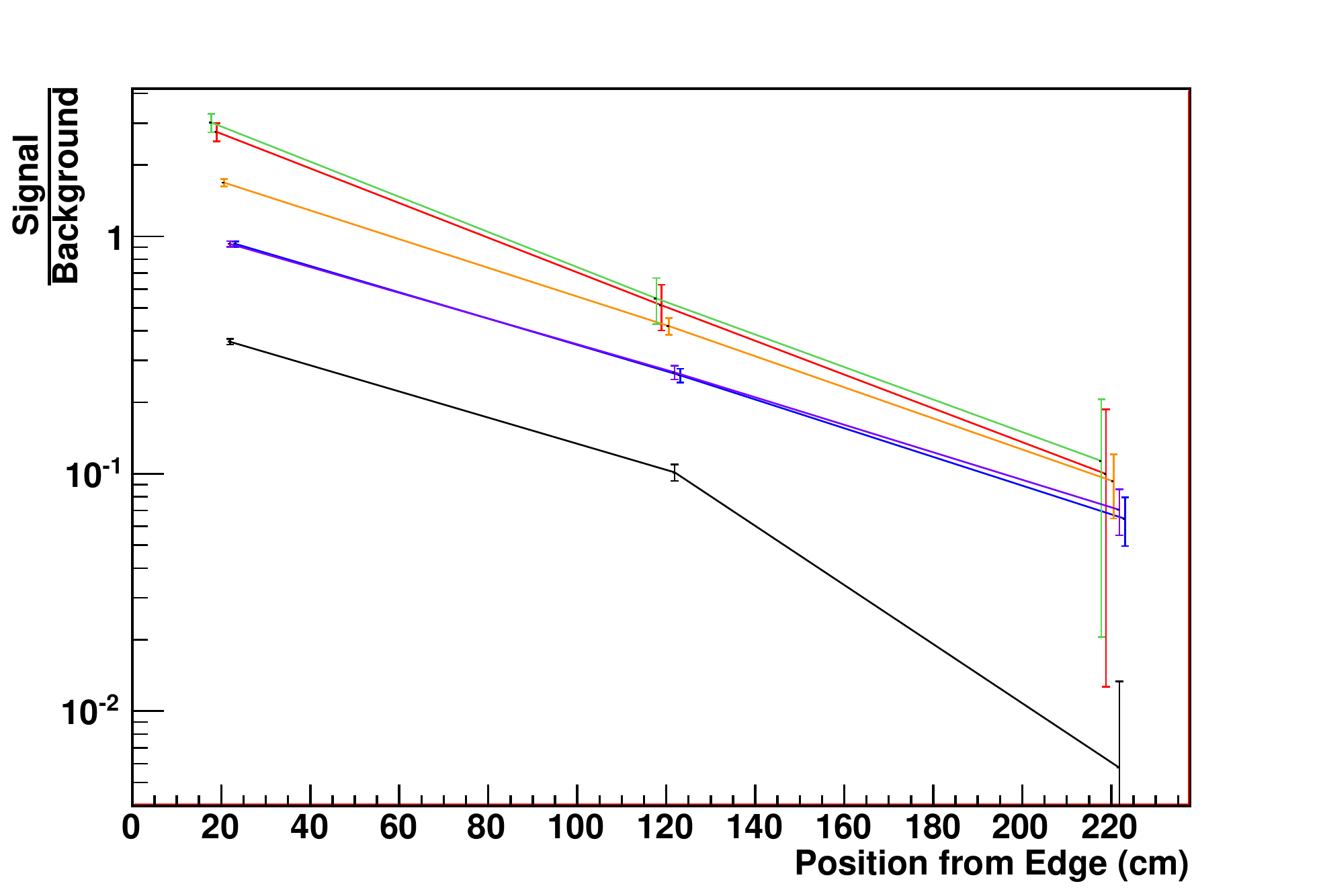}
\caption{The signal-over-background ratio for $^{252}$Cf adjacent to the detector, one meter from the detector's edge, and two meters from the detector edge.  The colors correspond to no cuts (black), the charge cut on the current event (blue), the current charge and muon veto cut (purple), the charge cut on the current and previous event (orange) resulting in neutron-neutron pairs, the neutron-neutron pairs plus the timing cut (red), and the neutron-neutron pairs plus the timing cut with the muon veto (green).  There is an arbitrary offset in the x-axis as a visual aid: the green curve corresponds to the true radial position.}
\label{fig:SOB}
\end{figure}

Based on the average number of neutrons produced from our $^{252}$Cf source, as well as the average number of neutrons produced in reactor grade plutonium (RGP), we have determined that our current setup can detect approximately $150~$g equivalent of RGP over background at two meter standoff from the detector edge in 20 seconds.  The solid angle presented by our detector at this distance is 2.4\% of 4$\pi$.   We have assumed 8\% fraction of $^{240}$Pu in RGP (87,000 n/s/kg).  Figure \ref{fig:SOB} is the signal over background for each cut in Table \ref{tab:cuts} as a function of radial distance from the detector.  At one meter from the detector edge, or 6.4\% of 4$\pi$, we have a signal to background ratio of 0.547 $\pm$ 0.120 after all cuts have been made.  Before cuts, the signal over background was 0.101 $\pm$ 0.00810.  At two meters, the signal over background before cuts is 0.00560 $\pm$ 0.00755 and increases to 0.113 $\pm$ 0.0927 after all cuts have been made.  Since our detector is intended as a correlated event detector, it is expected that high solid angle coverage is required for efficient operation.  However, with the proper event-level cuts our source is detectable even at two meters away: without requiring neutron-neutron pairs, the signal over background is 0.0702 $\pm$ 0.0154.

\subsection{Next Generation Detector}
There are many ways to improve the performance of this detection technology.  Aside from optical and geometrical optimizations, decreasing the neutron capture window by increasing the concentration of GdCl$_3$ will result in a decrease in the rate of accidental coincidences.  An examination of this detector's $^{252}$Cf calibration data shows that if the neutron capture window was decreased from 35 $\mu$s to 10 $\mu$s, and assuming the same number of neutrons are captured, the percentage of accidental coincidences drops from 53\% of all events surviving the current and last event energy cuts to 23\%.  In other words, if all neutrons are captured within 10 $\mu$s, then for every 4.3 coincident events, one would be accidental.  This is a significant improvement compared to our current concentration in which all neutrons are captured within 35$\mu$s: for every 1.9 coincident events, one is accidental.  It has been shown that there is no measurable impact on water attenuation length at the level of 0.2\% GdCl$_3$ \cite{coleman}.  An increase in GdCl$_3$ concentration to the 0.3\% level is needed to obtain a characteristic capture time of $\sim$10 $\mu$s.  Further study may be warranted in this area. 

\subsection{Conclusions}
We have successfully operated a large-scale gadolinium-doped water \v{C}erenkov detector and characterized its neutron detection performance for the purposes of SNM monitoring. Using a tagged americium beryllium source, the raw neutron detection efficiency has been measured in the center of the detector at 70\% and outside the detector at 31\%. Event-level cuts have been established to maximize the detection of correlated pairs of neutrons emitted simultaneously from a $^{252}$Cf fission source.  We have demonstrated detection of approximately $150~$g equivalent of RGP over background within 20 seconds for a solid angle coverage of 2.4\% of 4$\pi$; an optimized geometry is expected to perform even better. 

The neutron capture response of the detector has been reproduced in Geant4 for two different source positions, using the water attenuation length, the PMT quantum efficiency, and the wall reflectivity as tuning parameters. The tuned value of each parameter is within the expected range for the water quality, PMTs, and Tyvek reflectivity. Future work will use this model to design an optimized radiation portal monitoring system for detection of correlated neutrons from undeclared fission sources within cargo containers.

\section*{Acknowledgements}
The authors would like to thank Dennis Carr for assistance with detector design and construction, as well as Serge Ouedraogo for help with construction.  The authors also wish to thank the DOE NA-22 for their support of this project.

This work was performed under the auspices of the U.S. Department of Energy by Lawrence Livermore National Laboratory under Contract DE-AC52-07NA27344.   Document release number LLNL-JRNL-479935.


\begin{thebibliography}{1}

\bibitem{geant4} J. Allison {\it et. al.} ``Geant4 developments and applications" {\it IEEE Transactions on Nuclear Science} {\bf 53(1)} (2006) p. 270.

\bibitem{sno} The SNO Collaboration. ``Electron energy spectra, fluxes, and day-night asymmetries of $^8$B solar neutrinos from measurements with NaCl dissolved in the heavy-water detector at the Sudbury Neutrino Observatory" {\it Physical Review C} {\bf 72} (2005) 055502.

\bibitem{SK} The Super-Kamiokande Collaboration. ``Solar neutrino measurements in Super-Kamiokande-I" {\it Physical Review D} {\bf 73} (2006) 112001.

\bibitem{coleman} W. Coleman, A. Bernstein, S. Dazeley and R. Svoboda. ``Transparency of 0.2\% GdCl3 doped water in a stainless steel test environment"  { \it Nuclear Instruments and Methods A} {\bf 595} (2008) p. 339.

\bibitem{dazeley} S. Dazeley, A. Bernstein, N.S. Bowden, and R. Svoboda. ``Observation of neutrons with a Gadolinium doped water Cherenkov detector" { \it Nuclear Instruments and Methods A} {\bf 607} (2009) p. 616.

\bibitem{apollonio}  M. Apollonio {\it et. al.}  ``Search for neutrino oscillations on a long base-line at the CHOOZ nuclear power station" {\it European Physical Journal C} {\bf 27} (2003) p. 331.

\bibitem{boehm} F. Boehm {\it et. al.}  ``Final results from the Palo Verde neutrino oscillation experiment" {\it Physical Review D} {\bf 64} (2001) 112001.

\bibitem{piepke} A. B. Piepke, S.W. Moser, and V.M. Novikov. ``Development of a Gd loaded liquid scintillator for electron anti-neutrino spectroscopy" {\it Nuclear Instruments and Methods A} {\bf432} (1999) p. 392.

\bibitem{quickenden} T. I. Quickenden and J. A. Irvin. ``The ultraviolet absorption spectrum of liquid water"  {\it Journal of Chemical Physics}  {\bf 72 (8)} (1980) p. 4416.

\bibitem{sogandares} F. M. Sogandares and E. S. Fry.  ``Absorption spectrum (340 - 640 nm) of pure water. I. Photothermal measurements" {\it Applied Optics} {\bf 36} (1997) p. 8699.

\bibitem{pope} R. M. Pope and E. S. Fry. ``Absorption spectrum (380 - 700 nm) of pure water. II. Integrating cavity measurements" {\it Applied Optics} {\bf 36} (1997) p. 8710.


\end{thebibliography}
\end{document}